\scrollmode
\documentclass{article}
\usepackage{epsfig}

\let\rho\varrho

\begin{document}

\title{Infinite compressibility states in the
Hierarchical Reference Theory of fluids.\\
II.~Numerical evidence}

\author{Albert~Reiner${}^{1,2,3}$ and Gerhard~Kahl${}^1$\\
${}^1$Institut f\"ur Theoretische Physik and\\
Center for Computational Materials Science,\\
Technische Universit\"at Wien,\\
Wiedner Hauptstra\ss e~8--10, A--1040 Vienna, Austria.\\
${}^2$Teoretisk fysikk, Institutt for fysikk,\\
Norges teknisk-naturvitenskapelige universitet Trondheim,\\
H\o gskoleringen~5, N--7491 Trondheim, Norway.\\
${}^3$e-mail: {\tt areiner@tph.tuwien.ac.at}.}

\maketitle

\begin{abstract}
Continuing our investigation into the Hierarchical Reference Theory of
fluids for thermodynamic states of infinite isothermal compressibility
$\kappa_T$ we now turn to the available numerical evidence to
elucidate the character of the partial
differential equation: Of the three scenarios identified previously,
only the assumption of the equations
turning stiff when building up the divergence of $\kappa_T$ allows for
a satisfactory interpretation of the data. In addition to the
asymptotic regime where the arguments of part~I\ directly apply, a
similar mechanism is identified that gives rise to transient stiffness
at intermediate cutoff for low enough temperature. Heuristic arguments
point to a connection between the form of the Fourier transform of the
perturbational part of the interaction potential and the cutoff where
finite difference approximations of the differential equation cease to
be applicable, and they highlight the rather special standing of the
hard-core Yukawa potential as regards the severity of the
computational difficulties.
\end{abstract}


\section{Introduction}

Building upon the results of part~I \cite{ar:10}, {\it q.~v.{}}, and
maintaining the
notational and semantic conventions introduced there, we now turn to
the numerical solution of the {\sc hrt}\ {\sc pde}\
\cite{b:hrt:1,hrt:1,hrt:2,hrt:3,hrt:4,hrt:8,hrt:10}
\begin{equation} \label{pde:f}
{\partial f\over\partial Q}
=  d_{00} +  d_{02}\,{\partial^2f\over\partial\rho^2}
\end{equation}
in order to determine the type of behavior that actually occurs in
practical applications of the theory for thermodynamic states of
diverging isothermal compressibility $\kappa_T$.  To this end we
consider two simple model potentials $v(r) = v^{{\rm hs}}(r) + w(r)$, {\it
viz.{}},
the hard-core Yukawa ({\sc hcy}) system,
\begin{equation} \label{pot:hcy}
w^{{\rm hcy}}(r) = \left\{\begin{array}{ccc}
-\epsilon_0 &:& r<\sigma \\
-\epsilon\,{\sigma\over r}\, e^{-z\,(r-\sigma)} &:& r>\sigma,
\end{array}\right.
\end{equation}
and square wells ({\sc sw}s),
\begin{equation} \label{pot:sw}
w^{{\rm sw}}(r) = \left\{\begin{array}{ccc}
-\epsilon &:& r<\lambda\,\sigma \\
0        &:& r>\lambda\,\sigma,
\end{array}\right.
\end{equation}
to illustrate the types of behavior encountered and to test the
predictions furnished by the relevant scenarios.  In both of these
potentials  $\epsilon$ coincides with the negative
of the contact value of the interaction, $\lim_{r \to \sigma+}
\left(-w(r)\right)$, and so sets the energy scale of the problem.  The
potential range is given by $1/z$ and $\lambda\,\sigma$,
respectively.  Unless stated otherwise,
$\epsilon_0$, the value of $w^{{\rm hcy}}(r)$ inside the core, coincides
with
$\epsilon$, a choice shared with the implementation by the authors of
{\sc hrt}\ and their coworkers referred to as the original one in
refs.~\cite{ar:4,ar:th}, {\it q.~v.{}}.  A short summary of the parameter
sets and
sample isotherms considered in this study can be found in
tab.~\ref{tab:params}.  In the numerical work we employ an unconditionally
stable implicit predictor-corrector scheme shortly characterized in
section~\ref{sec:numerics}.  A more extensive discussion of the
implementation can be found in refs.~\cite{ar:4,ar:th}, where default
settings for the most important customization parameters are also
documented.  Even further technical information is available with the
source distribution itself \cite{ar:hrt:1}.

\begin{table}
\begin{tabular}{|lll|ccc|cccc|}
&system&&$\beta_c\,\epsilon$&$\rho_c\,\sigma^3$&&$\beta\,\epsilon$&$
\rho_v\,\sigma^3$&$ \rho_l\,\sigma^3$&\\
\hline
&{\sc sw}, $\lambda=3$&&0.1011&0.26(1)&&0.115&0.075(5)&0.510(5)&\\
&{\sc hcy}, $z=1.8/\sigma$&&0.8316&0.33(1)&&0.875&0.145(5)&0.525(5)&\\
\end{tabular}
\caption{Overview of systems and sample isotherms: $\beta_c$ and $\rho_c$
give the location of the critical point, $ \rho_v$ and $ \rho_l$ the
extent of the two-phase region at the inverse temperature $\beta$
considered in the tables and figures to follow.  The numbers have
been obtained from {\sc hrt}\ calculations not imposing the core
condition.  All of the digits indicated for $\beta_c$ are
significant.}\label{tab:params}
\end{table}

Of the three types of behavior compatible with the local properties of
the {\sc pde}, both genuine ($r = s = 0$) and effective ($r > 0$, $s = r +
1 > 1$, $ r_{\rm eff} =  s_{\rm eff} = 0$) smoothness imply a {\sc fd}\
approximation to
$f$ growing like $1/Q$.  The monotonous scenario, on the other hand,
furnishes the
specific prediction that $\bar\varepsilon\,Q^2$ tends to a finite limit for
$Q\to0$.  As we will see in section~\ref{sec:monotonicity:refuted}, the
numerical evidence clearly rules out this possibility.

It is thus only the genuinely smooth and the stiff scenarios that
remain to be considered in section~\ref{sec:decision}.  The results of the
computations reported there do, indeed, allow us to infer the
character of the {\sc pde}\ for subcritical temperatures, $T <  T_c$, with
great confidence, if only indirectly due to the great computational
similarity of genuine and effective smoothness.  Our main evidence in
favor of the stiff scenario derives from the rather detailed and
testable predictions it entails, all of which are confirmed
numerically.  By way of contrast, the genuinely smooth scenario does
not hold an explanation for the observed trends, especially as regards
the dependence of the {\sc fd}\ results on the properties of the
discretization grids.

Our conclusion that the {\sc pde}\ actually turns stiff in part of ${\cal
D}$ for
$T \le  T_c$ then paves the way for some heuristic arguments relating
the onset of smoothing in $Q$ to the form of
the Fourier transform of the perturbational part of the potential
(section~\ref{sec:heuristics}).  So having understood the behavior of the
{\sc pde}\ in the limit $Q\to0$ where asymptotic reasoning valid for large
$\bar\varepsilon$ applies, in section~\ref{sec:beyond} we then turn to
similar
computationally problematic features of its solution at much higher
cutoff where the numerical evidence points to a mechanism not unlike
that at work in the asymptotic region.  We close with an informal
discussion of the reasons for the atypical computational properties of
{\sc hcy}\ fluids of moderate inverse screening length $z$
(section~\ref{sec:hcy}).

\section{The monotonicity assumption refuted}

\label{sec:monotonicity:refuted}

\begin{figure}[t]\vbox{\noindent\epsfbox{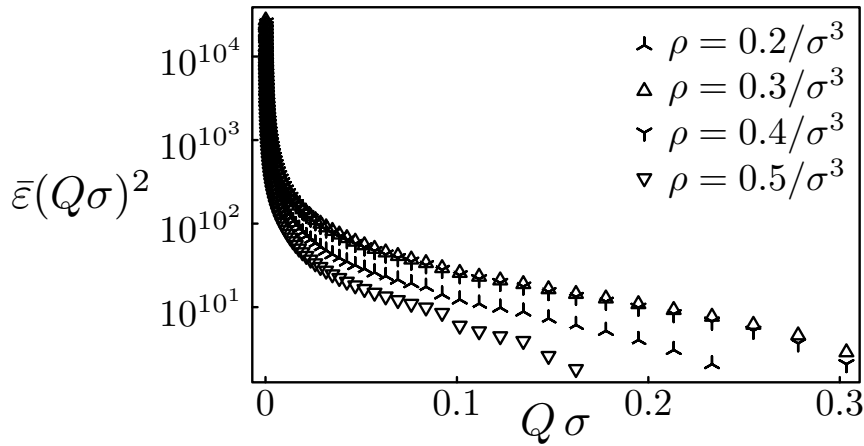}
\bigskip\caption{$\bar\varepsilon\,Q^2$ as a function of $Q$ for
various densities inside the binodal: The data have been obtained
for a hard-core Yukawa potential with inverse screening length
$z=1.8/\sigma$ and for an inverse temperature of
$\beta=0.875/\epsilon$.  The numerical precision in the
calculations was ${\epsilon_{\tt\#}}=10^{-2}$, the step size for infinite
cutoff was ${\Delta
Q\big\vert_\infty}=10^{-2}/\sigma$.}\label{fig:epsbarQ2}}
\end{figure}

According to section~V of part~I \cite{ar:10}, the assumption
of a merely logarithmic
divergence of $f$ furnishes the rather specific prediction
of $\bar\varepsilon\,Q^2$ tending to a finite limit for $Q\to0$. Of course,
the possiblity of non-zero $s$ means that, in principle, the smoothing
effect discussed
in section~VI\ of part~I \cite{ar:10}\ must be reckoned with.  The
singularity being so mild, however, a possible reduction of $s > 0$ to
an effective value of $ s_{\rm eff} = 0$ is preempted by the choice of step
sizes $\Delta Q$:

In our implementation of the theory the cutoff in the $i^{\rm th}$
{\sc fd}\ step is parametrized as
\begin{displaymath}
Q_{(i)}=\ln\left(e^{a-i\,b}+1\right)/\sigma,\qquad i=0,1,\ldots,
\end{displaymath}
as is the case for the program the original authors of {\sc hrt}\ and their
coworkers employ, too.  Here $a/\sigma$ is close to the cutoff $Q_{(0)}
\equiv {Q_{\infty}}$ where initial conditions are imposed on $f$,
and $b/\sigma$ is the spacing ${\Delta Q\big\vert_\infty}$ of successive
cutoffs
in the large $Q$ limit.   For $Q \to 0$, on the other hand, we easily find
$\Delta Q \approx -Q \,
\left(1-e^{-b}\right)$. If $\bar\varepsilon\,Q^2$ is to approach zero or a
finite constant as predicted by the assumption of monotonous growth
these step sizes thus turn out of order ${\bf O}(\bar\varepsilon^{-1/2})$
at most,
and our discretization should allow us to follow the variation of $f$
reasonably well all the way to $Q=0$. From
fig.~\ref{fig:epsbarQ2}, however, we see that $\bar\varepsilon\,Q^2$
clearly
diverges for $Q \to 0$.  As this finding is
corroborated by further calculations with a smaller setting of the
numerical parameter ${\Delta Q\big\vert_\infty}$, on finer density grids
(down to
$\Delta\rho = 5\cdot10^{-4} / \sigma^3$), and for both hard-core
Yukawa and square well potentials we feel we can safely exclude the
monotonous growth scenario from further consideration.

\section{Smoothness {\it vs.{}} stiffness}

\label{sec:decision}

\begin{figure}[t]\vbox{\noindent\epsfbox{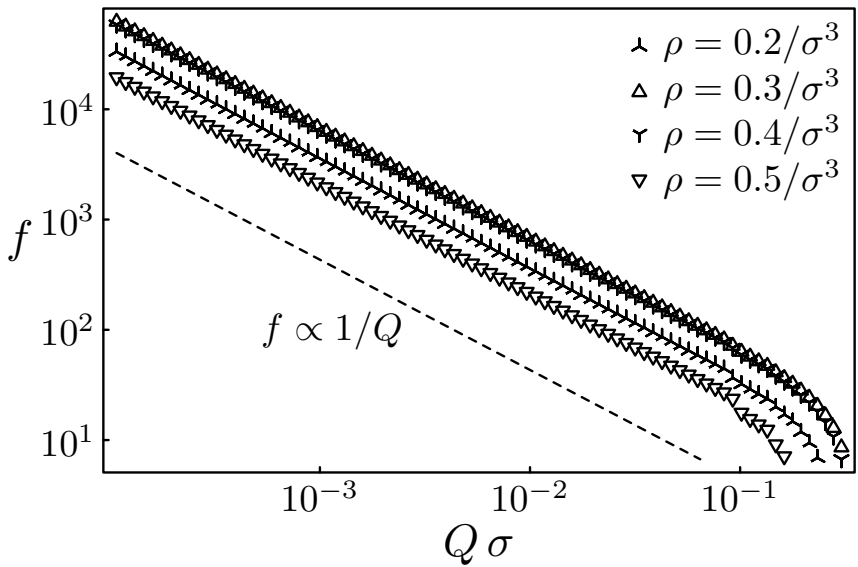}
\bigskip\caption{$f$ as a function of $Q$ for various
densities inside the binodal: The data have been obtained for the
same hard-core Yukawa potential and with the same numerical
parameters as in fig.~\ref{fig:epsbarQ2}.  The dashed line indicates
the slope corresponding to proportionality of $f$ to the
reciprocal of the cutoff.  Subsequent symbols are separated by ten
steps in the $-Q$ direction.}\label{fig:f:hcy}}
\end{figure}

As for the remaining two alternatives, an attempt to distinguish
numerically between genuine and effective smoothness seems doomed at
first sight: both predict a numerically smooth solution growing like
$1/Q$ and with a profile like that sketched in fig.~1 of
part~I \cite{ar:10}.  And indeed, fig.~\ref{fig:f:hcy} shows the small $Q$
behavior of
$f$ within the binodal as obtained numerically to be in excellent
agreement with $f\propto1/Q$, and figs.~\ref{fig:f:hcy}, \ref{fig:f:sw},
and~\ref{fig:f:contour} as well as the numerical data demonstrate that $f$
is of the form necessary for a stable pattern of growth, as postulated
in part~I \cite{ar:10}, {\it q.~v.{}}.  But even though these general
features fit both
scenarios, close scrutiny of the computational process and the
numerical results yields a wealth of indirect evidence that we feel is
sufficient to establish the stiffness of the {\sc pde}\ for $T <  T_c$ with
great confidence, if not with absolute certainty.  We base our
reasoning on the rather specific, and numerically testable predictions
that follow from the assumption of stiffness.  These stand in marked
contrast to the vague expectations furnished by the genuinely smooth
scenario that can, however, never be ruled out completely as
smoothness is its sole defining characteristic.  As we will show in
this section, it is the assumption of a stiff {\sc pde}\ that is in full
accordance with the numerical findings whereas smoothness is only
marginally compatible with some of their traits, especially as regards
the {\sc sw}\ data.

Before, however, some general remarks are in place: Letting the labels
$x$ and $y$ refer to either $Q$ or $\rho$, in the stiff scenario
smoothing in $x$ sets in at $Q={Q_{\Delta x}}$ and can always be postponed,
{\it i.~e.{}}, shifted to lower cutoffs by decreasing the step size $\Delta
x$.
If, however, the corresponding exponent, $r$ or $s$, is positive, the
rapid growth of $f$, $\bar\varepsilon$ and $\vert\partial^2f/\partial
x^2\vert$ as the solution
proceeds towards $Q=0$ implies that the amount by which ${Q_{\Delta x}}$
can
be changed in this way and the attendant computational effects must be
small.  For positive exponents the ${Q_{\Delta x}}$ are thus fairly well
defined despite the gradual nature of the transition to the smoothing
regime.  --- Furthermore, without loss of generality assuming ${Q_{\Delta
x}}
> {Q_{\Delta y}}$, ${Q_{\Delta x}}$ is obviously independent of the step
sizes
$\Delta y$.  The solution obtained numerically at cutoffs below
${Q_{\rm smooth}} \equiv {Q_{\Delta x}}$ is already affected by smoothing
in $x$ so
that there is no point in identifying ${Q_{\Delta y}}$ with the cutoff
where
$\Delta y$ becomes too large to describe the variation of the no
longer accessible true solution of the {\sc pde}.  Instead, ${Q_{\Delta
y}}$ is
taken to be the cutoff where smoothing in $y$ commences in the
solution of the {\sc fd}\ equations ({\sc fde}s), which implies a $\Delta
x$
dependence of ${Q_{\Delta y}}$ and may even induce ${Q_{\Delta y}}$ to
vanish
altogether. For $Q < {Q_{\Delta y}}$, the solution generated numerically by
necessity conforms to the smooth scenario as $ r_{\rm eff} =  s_{\rm eff} =
0$ and
so grows like $1/Q$ in a stable manner.  This proportionality also
means that the form of $f$ remains
constant from ${Q_{\Delta y}}$ all the way to the final results at zero
cutoff.  (Here and below the form of $f$ at some cutoff $Q$ refers to
$f(Q, \rho)$ as a function of $\rho$, restricted to $\rho_1 < \rho <
\rho_2$ and without regard for the overall normalization of $f$.)  ---
The important mechanism sketched in section~III\ of part~I \cite{ar:10}\
and the concomitant stabilization of form and monotonicity of $f$ do
not explicitly depend on $s$ and thus always set in at ${Q_{\Delta \rho}}$;
incidentally, figs.~\ref{fig:f:hcy} and~\ref{fig:f:sw} show its
preconditions,
{\it viz.{}}, flatness and compatibility with the sketch of part~I
\cite{ar:10}\ to be met
numerically.  Of course, both ${Q_{\Delta Q}}$ and ${Q_{\Delta \rho}}$
depend on
temperature and density, which is taken to be silently accounted for
whenever we speak of the form of $f$ at one of the ${Q_{\Delta x}}$, and
they
are defined only in that part of ${\cal D}$ where $f$ is large.  --- Not
surprisingly, the two possible orderings for the cutoffs ${Q_{\Delta Q}}$
and
${Q_{\Delta \rho}}$ assigned in an interpretation of the numerical results
in
terms of the stiff scenario entail vastly different consequences and
are therefore discussed separately in subsections~\ref{sec:smooth:rho:Q}
(${Q_{\Delta \rho}} > {Q_{\Delta Q}}$) and~\ref{sec:smooth:Q:rho}
(${Q_{\Delta Q}} > {Q_{\Delta \rho}}$) below.

Before that, however, it is worthwhile to step back for a moment and
ask why we have to adopt eq.~(\ref{pde:f}) in the first place if the most
direct formulation of the theory is that of a {\sc pde}\ for the free
energy $A^{(Q)}(\rho)$ of the $Q$ system at density $\rho$, {\it cf.{}}\
part~I\
\cite{ar:10}.  Indeed, from eqs.~(A2) and~(A3) of part~I\
\cite{ar:10} we see that $\partial A^{(Q)} / \partial Q \propto Q^2 \, (f +
{\rm const})$ for $Q\,\sigma \ll 1$ so that the $Q$ and $\rho$ scales
characteristic of $A^{(Q)}(\rho)$ are essentially the same as for $f(Q,
\rho)$. In the smooth scenario there is then
no reason for the formulation in terms of $f$ to be preferable to that
in terms of the free energy, provided proper care is taken to ensure
stability and convergence. This has certainly been the case in our
earlier work shortly summarized in appendix~B.1 of ref.~\cite{ar:th} that
nevertheless was unable to proceed to small $Q$ for $T <  T_c$. Similar
difficulties are reported in ref.~\cite{hrt:4}, and to the best of our
knowledge there are no {\sc hrt}\ results on simple one-component fluids
for $T <  T_c$ except in the quasilinear formulation of eq.~(\ref{pde:f})
or
variants thereof.  --- In the stiff scenario all this is, of course,
to be expected as the rapid low amplitude oscillations of the solution
in this case necessitate step sizes that are reduced as some inverse
power of $\bar\varepsilon$ or the exponential of $\partial A^{(Q)} /
\partial Q$,
and only under special circumstances do the discretized equations
allow one to obtain a solution with the much larger step sizes used in
practical applications.  As noted in section~II\ of part~I \cite{ar:10},
the auxiliary quantity $f(Q, \rho)$ was introduced exactly for this
reason \cite{hrt:10}.

\subsection{Numerical aspects}

\label{sec:numerics}

As some of the numerical effects are rather subtle, we should also
recall several key aspects of the implementation we rely on.  This is
a highly flexible and fully modular computational framework for the
solution of a {\sc fd}\ approximation of the {\sc pde}\ by an implicit
predictor-corrector scheme thoroughly discussed in refs.~\cite{ar:4,ar:th}.
For consistency with part~I \cite{ar:10}, in the calculations reported here
we refrain from implementing the core condition.  The discretization
is applied on uniform density grids and with the predetermined step
sizes $\Delta Q$ of section~\ref{sec:monotonicity:refuted}.  Convergence of
the {\sc fd}\ equations has been checked, and iteration of the corrector
step does not bring about noticeable changes.

In practical applications, the discretized equations generally cannot
be solved down to arbitrarily small $Q$ for $T <  T_c$, and the
smallest cutoff reached we denote ${Q_{\rm min}}$. As the failure modes
responsible for an end of the program are known \cite{ar:4,ar:th} and
can be linked to the local behavior of the solution, {\it v.~i.{}}, the
systematic changes in ${Q_{\rm min}}$ upon variation of aspects of the
numerical procedure provide a powerful and readily accessible
diagnostic tool.  For the calculations reported here, the immediate
cause for abortion of the computation at some cutoff ${Q_{\rm min}}$ is
either
an insufficient adaptation of the rescaling necessary for representing
quantities affected by exponentiation of $f$ --- the scale of
fig.~\ref{fig:epsbarQ2}\ alone shows that, {\it e.~g.{}}, $\bar\varepsilon$
cannot be
represented in double precision --- or else because of non-real $f$
and negative $\varepsilon \equiv \bar\varepsilon + 1$ in the predictor
step.  These
two effects are linked to rapid increase and decrease of $f$, respectively.

Unlike the ${Q_{\Delta x}}$, ${Q_{\rm min}}$ obviously does not depend on
the
density. Instead, it is essentially determined by the physical
potential $w(r)$, the temperature, the discretization grid, and the
formulation of the theory \cite{ar:4}. As for the latter, if the {\sc pde}\
is coupled to further constraints, and the solution vector augmented
by additional components to be determined accordingly, the likelihood
of an early termination of the computation in the predictor step
generally increases, and so does ${Q_{\rm min}}$.  As the customary manner
of
implementing the core condition involves an expansion of the direct
correlation function inside the core \cite{hrt:4,ar:4}, the sensitivity
of ${Q_{\rm min}}$ to an increase in ${N_{\rm cc}}$, the number of
expansion
coefficients, again proves of interest.

For a more detailed account we refer the reader to refs.~\cite{ar:4,ar:th}
as well as the documentation that comes with the source code
distribution itself \cite{ar:hrt:1}.

\subsection{Smoothing in $\rho$ first}

\label{sec:smooth:rho:Q}

So let us first turn to the {\sc hcy}\ fluid of inverse screening length $z
= 1.8/\sigma$ already considered in ref.~\cite{ar:4}.  As mentioned
before, the numerical solution must be smooth at any rate and is
therefore always compatible with the genuinely smooth scenario.  In
this case we expect only a small dependence of the results on $\Delta
Q$ and $\Delta\rho$ that should be essentially stochastic in nature,
stemming from the truncation error in an otherwise unproblematic {\sc fd}\
approximation of the {\sc pde}\ alone.

As we shall see in a moment, the numerics can also be reconciled with
the stiff scenario if only we assume smoothing to occur in the $\rho$
direction first, ${Q_{\Delta \rho}} > {Q_{\Delta Q}}$, furnishing the
following
predictions: The mechanism responsible for stable growth of $f$ ({\it
cf.{}}\
section~III\ of part~I \cite{ar:10}) being at work at all cutoffs below
${Q_{\rm smooth}}$, the stability of the computational process is not an
issue
and incorporation of the core condition is entirely unproblematic.  An
overflow due to an insufficient adaptation of the re-scaling of
non-${\bf O}(1)$ quantities is the only possibility for numerical
failure, and its likelihood is greatly reduced when $\Delta Q$ is
decreased so that smaller step sizes are generally accompanied by
smaller values of ${Q_{\rm min}}$.  A systematic $\Delta\rho$ dependence of
${Q_{\rm min}}$ is not anticipated.  --- For fixed density grid,
${Q_{\Delta \rho}} \equiv {Q_{\rm smooth}}$ cannot depend on $\Delta Q$,
nor can $f$
at ${Q_{\Delta \rho}}$.  On the other hand, even though smaller step sizes,
{\it i.~e.{}}, smaller ${\Delta Q\big\vert_\infty}$,
{\it cf.{}}\ section~\ref{sec:monotonicity:refuted}, correspond to smaller
${Q_{\Delta Q}}$,
$s>1$ implies that the drop in ${Q_{\Delta Q}}$ must be exceedingly small. 
As
furthermore the evolution from ${Q_{\Delta \rho}}$ down to ${Q_{\Delta Q}}$
is determined
by the solution at the onset of smoothing and the properties of only
the density grid, the form of $f$ below ${Q_{\Delta Q}}$, including $Q=0$,
is
virtually $\Delta Q$ independent.  --- As for a variation of the
density grid at fixed $\Delta Q$, a reduction of $\Delta\rho$ clearly
entails a shift of ${Q_{\rm smooth}} \equiv {Q_{\Delta \rho}}$ to smaller
cutoffs, which
may in turn cause a change in ${Q_{\Delta Q}}$, too.  These effects must be
rather small because of the non-zero exponenets $r$ and $s$, and they
must vary with the density for the same reason the ${Q_{\Delta x}}$ are
density dependent.  A change of the $\rho$ grid thus implies a small
change of the form of $f$ at ${Q_{\Delta \rho}}$ and, hence, at ${Q_{\Delta
Q}}$ and all
smaller cutoffs.  As long as ${Q_{\Delta \rho}}$ does not fall below
${Q_{\Delta Q}}$,
however, the ratio of the forms of $f$ as obtained on different
density grids cannot depend on $\Delta Q$.

\begin{table}
\begin{tabular}{|c|cc|ccc|cccc|}
${\Delta Q\big\vert_\infty}\,\sigma$&${Q_{\rm
min}}\,\sigma$&&$f_{0.2}$&$f_{0.2}\,{Q_{\rm
min}}\,\sigma$&&$F{}_{0.2}^{0.3}$&$F{}_{0.2}^{0.4}$&$F{}_{0.2}^{0.5}$&\\
\hline
0.003&$9.914\cdot10^{-3}$&&$3.643\cdot10^2$&$3.612$&&1.862&1.755&0.590&\\
0.004&$3.995\cdot10^{-5}$&&$8.990\cdot10^4$&$3.592$&&1.867&1.758&0.585&\\
0.005&$5.014\cdot10^{-5}$&&$7.163\cdot10^4$&$3.592$&&1.867&1.758&0.585&\\
0.010&$9.943\cdot10^{-5}$&&$3.612\cdot10^4$&$3.592$&&1.867&1.758&0.585&\\
\end{tabular}
\caption{$\Delta Q$ dependence of the final results for a hard-core Yukawa
system: Just as in figs.~\ref{fig:epsbarQ2} and \ref{fig:f:hcy},
$z=1.8/\sigma$, $\beta=0.875/\epsilon$, and
$\Delta\rho=10^{-2}/\sigma^3$.  We use the notation $f_x$ for
$f({Q_{\rm min}}, x/\sigma^3)$ and define $F{}^y_x \equiv
f_y/f_x$.}\label{tab:Qdrho:hcy:100}
\end{table}

\begin{table}
\begin{tabular}{|c|cc|ccc|cccc|}
${\Delta Q\big\vert_\infty}\,\sigma$&${Q_{\rm
min}}\,\sigma$&&$f_{0.2}$&$f_{0.2}\,{Q_{\rm
min}}\,\sigma$&&$F{}_{0.2}^{0.3}$&$F{}_{0.2}^{0.4}$&$F{}_{0.2}^{0.5}$&\\
\hline
0.003&$3.131\cdot10^{-5}$&&$1.167\cdot10^5$&$3.652$&&1.865&1.768&0.625&\\
0.004&$1.004\cdot10^{-2}$&&$3.638\cdot10^2$&$3.653$&&1.865&1.768&0.625&\\
0.005&$5.014\cdot10^{-5}$&&$7.285\cdot10^4$&$3.652$&&1.865&1.768&0.625&\\
0.010&$1.004\cdot10^{-4}$&&$3.637\cdot10^4$&$3.653$&&1.865&1.768&0.625&\\
\end{tabular}
\caption{$\Delta Q$ dependence of the final results for a hard-core Yukawa
system: The parameters and notation coincide with those of
tab.~\ref{tab:Qdrho:hcy:100}, except for $\Delta\rho =
5\cdot10^{-4}/\sigma^3$.}\label{tab:Qdrho:hcy:2000}
\end{table}

\begin{table}
\begin{tabular}{|c|ccccc|}
${\Delta
Q\big\vert_\infty}\,\sigma$&$G_{0.2}$&$G_{0.3}$&$G_{0.4}$&$G_{0.5}$&\\
\hline
0.003&1.011&1.013&1.018&1.072&\\
0.004&1.017&1.016&1.022&1.087&\\
0.005&1.017&1.016&1.022&1.086&\\
0.010&1.017&1.016&1.022&1.086&\\
\end{tabular}
\caption{$\Delta\rho$ dependence of the form of the final results for a
hard-core Yukawa system at varying $\Delta Q$: The parameters
coincide with those of tabs.~\ref{tab:Qdrho:hcy:100}
and~\ref{tab:Qdrho:hcy:2000}.  Perusing the notation introduced there,
$G_x$ is $f_x\,{Q_{\rm min}}\,\sigma$ as evaluated for
$\Delta\rho=5\cdot10^{-4}/\sigma^3$ divided by the same quantity
for $\Delta\rho=10^{-2}/\sigma^3$.}\label{tab:Qdrho:hcy}
\end{table}

All these predictions are confirmed in the actual calculations for a
{\sc hcy}\ potential with $z = 1.8 / \sigma$ on density grids with
$\Delta\rho = 10^{-2} / \sigma^3$ and $\Delta\rho = 5\cdot10^{-4} /
\sigma^3$ and varying $\Delta Q$ as summarized in
tabs.~\ref{tab:Qdrho:hcy:100} to~\ref{tab:Qdrho:hcy} and
fig.~\ref{fig:f:hcy}. For fixed density grid (tabs.~\ref{tab:Qdrho:hcy:100}
and~\ref{tab:Qdrho:hcy:2000}, respectively), ${Q_{\rm min}}$ and, hence,
the final
values of $f$ markedly depend on $\Delta Q$, the former generally
decreasing and the latter increasing upon reduction of the step size.
On the other hand, both the form of $f$ and its magnitude at fixed cutoff
--- to
be found in the tables under the headings of $F{}^y_x$ and
$f_x\,{Q_{\rm min}}\,\sigma$, respectively --- remain largely unchanged.
Comparing the results obtained with different settings for
$\Delta\rho$, the change in the final values of $f$ is indeed almost
completely due to the differences in ${Q_{\rm min}}$.  The magnitude at
fixed
$Q$, on the other hand, is affected only moderately, {\it viz.{}}, by a few
per cent for a
twenty-fold increase in the density resolution, and it depends on
$\rho$ but not on $\Delta Q$, {\it cf.{}}\ tab.~\ref{tab:Qdrho:hcy}. 
${Q_{\rm min}}$
itself is not affected by the density grid in a systematic way.  There
are two sample isotherms, {\it viz.{}}, the ones at ${\Delta
Q\big\vert_\infty} = 0.003 /
\sigma^3$, $\Delta\rho = 10^{-2} / \sigma^3$ and at ${\Delta
Q\big\vert_\infty} =
0.004 / \sigma^3$, $\Delta\rho = 5\cdot10^{-4} / \sigma^3$, that
founder at comparatively large cutoff.  Of these, only the former does
not enter the asymptotic regime where $f \propto 1/Q$, as can clearly
be seen from tab.~\ref{tab:Qdrho:hcy}.  All in all, the numerical results
are in excellent agreement with stiffness, and we note that for this
system and the density grids considered ${Q_{\Delta Q}}$ must be sought
around
$10^{-2}/\sigma$.  Trivially, being smooth the results also conform to
the smooth scenario as mentioned before.

\subsection{Smoothing in $Q$ first}

\label{sec:smooth:Q:rho}

In our previous work on {\sc hrt}\ \cite{ar:4,ar:5} we repeatedly stressed
the vastly different numerical properties of the {\sc hcy}\ and {\sc sw}\
potentials.  This is certainly not anticipated for genuine smoothness
that merely predicts a small grid dependence stemming from the local
truncation error of the discretization, exactly as for the {\sc hcy}\
system.  Still, the assumption of a genuinely smooth solution is
certainly compatible with the numerics, if only marginally so in the
face of the most prominent feature of the evolution of $f$, {\it viz.{}},
episodes of much more rapid variation than mere proportionality to
$1/Q$.

Assuming the {\sc pde}\ to turn stiff for large $f$ instead, and
furthermore ${Q_{\Delta Q}}$ to exceed ${Q_{\Delta \rho}}$ for the present
system, there is
a $Q$ range ${Q_{\Delta \rho}}<Q<{Q_{\Delta Q}}$ where {\sc fde}s\ are used
with inappropriately
large step sizes $\Delta Q$ while oscillations in $\rho$ are not yet
suppressed. For these cutoffs, the stabilization wrought by the
mechanism analyzed in section~III\ of part~I \cite{ar:10}\ is not
effective yet, and there is no reason for $f$ to be convex from below
throughout the density range $\rho_1<\rho<\rho_2$.  On the other hand, the
overall profile of $f$
is expected to resemble fig.~1\
of part~I \cite{ar:10}, and $ s_{\rm eff}=0$ once more suggests a general
growth
proportional to $1/Q$.  The sign of $\partial^2f/\partial\rho^2$ is
thus unconstrained, and its modulus increases in unison with $f$, {\it
i.~e.{}},
in proportion to $1/Q$.  As $ d_{00} /  d_{02}$ is of order ${\bf O}(1)$,
however, the ${\bf O}(1)$ growth of $\partial^2 f / \partial \rho^2$ may
well be sufficient to destabilize the growth at some $Q \in ({Q_{\Delta
\rho}}, {Q_{\Delta Q}})$,
prompting much more rapid variation of $f$ as a function of $Q$.  Of
course, these near-discontinuities of $f$ will occur at different
cutoffs for different densities, most often close to
$\rho_1$ and $\rho_2$ where the
${Q_{\Delta x}}$ are smallest, and neighboring densities will experience
them
at roughly the same cutoff.  Furthermore, in principle the jumps
should lead to both increases and decreases in $f$, depending on the
sign of $\partial^2 f / \partial \rho^2$ at slightly larger $Q$.
Considering the numerics, however, a large change in $f$ is almost
certain to bring the calculation to an end, and all the failure modes
discussed in section~\ref{sec:numerics} are relevant for ${Q_{\rm min}}$. 
A
comparatively mild increase of $f$, on the other hand, may relax the
relative
curvature of $f$ to the point of allowing the solution to enter once
more an episode of near-stability characterized by growth in
approximate proportion to $1 / Q$.  As for an incorporation of the
core condition, in accordance with section~\ref{sec:numerics} the attendant
introduction of additional degrees of freedom is likely to exacerbate
the risk of triggering such a jump in $f$, {\it cf.{}}\
section~\ref{sec:beyond}.
--- To understand the grid dependence of the numerics under the
assumption of stiffness, recall that ${Q_{\rm min}}$ itself is the location
of a failed jump in $f$.  As smoothing in $Q$
is the driving force behind the computational process, ${Q_{\rm min}}$ must
be
quite sensitive to $\Delta Q$, but there is no reason for ${Q_{\rm min}}$
to
be monotonous in $\Delta Q$.  The density grid, on the other hand, is still
adequate for the elliptic boundary value problem in $\rho$ at constant
$Q$.  If the numerical process were stable, there should thus be no
appreciable dependence of the results on $\Delta \rho$ at all.  In the
absence of the stabilization wrought by smoothing in $\rho$, however,
even the small differences seen upon variation of $\Delta\rho$ must be
expected to shift the episodes of rapid evolution to slightly
different cutoffs in an unsystematic way.  By the same token, the
$\Delta \rho$ dependence of the final form of $f$ should be small, and
different $\Delta Q$ should leave it unaltered as long as the number
and the approximate positions of the jumps do not change.  As those
are least frequent close to the maximum of $f$, its form is expected
to be most stable in the central part of the density interval of large
$f$.

\begin{figure}[t]\vbox{\noindent\epsfbox{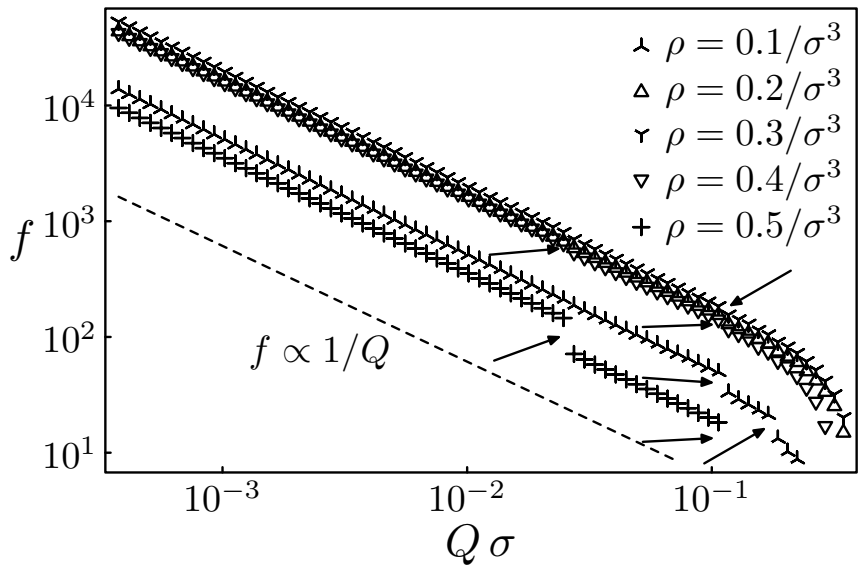}
\bigskip\caption{$f$ as a
function of $Q$ for various densities inside the binodal: The data
have been obtained for a {\sc sw}\ potential with $\lambda=3$ and at a
temperature of $\beta=0.115/\epsilon$; otherwise, the remarks of
fig.~\ref{fig:f:hcy} apply. Arrows mark several of the
near-discontinuities discussed in
section~\ref{sec:smooth:Q:rho}.}\label{fig:f:sw}}
\end{figure}

\begin{table}
\begin{tabular}{|c|cc|cc|cccc|}
${\Delta Q\big\vert_\infty}\,\sigma$&${Q_{\rm
 min}}\,\sigma$&&$f_{0.15}$&&$F{}_{0.15}^{0.25}$&$F{}_{0.25}^{0.35}$&$F{}_{0.35}^{0.45}$&\\
\hline
0.003&$4.181\cdot10^{-4}$&&$-1.415\cdot10^5$&&2.379&1.635&0.392&\\
0.004&$3.198\cdot10^{-4}$&&$\phantom{-}3.597\cdot10^4$&&1.589&0.965&0.523&\\
0.005&$3.318\cdot10^{-4}$&&$\phantom{-}3.465\cdot10^4$&&1.589&0.965&0.523&\\
0.010&$3.576\cdot10^{-4}$&&$\phantom{-}3.225\cdot10^4$&&1.589&0.965&0.523&\\
\end{tabular}
\caption{$\Delta Q$ dependence of the final results for a {\sc sw}\ system:
Just
as in fig.~\ref{fig:f:sw}, $\lambda=3$, $\beta=0.115/\epsilon$, and
$\Delta\rho = 10^{-2} / \sigma^3$.  The notation is the same as in
tab.~\ref{tab:Qdrho:hcy:100}.}\label{tab:Qdrho:sw3:100}
\end{table}

\begin{table}
\begin{tabular}{|c|cc|cc|cccc|}
${\Delta Q\big\vert_\infty}\,\sigma$&${Q_{\rm
 min}}\,\sigma$&&$f_{0.15}$&&$F{}_{0.15}^{0.25}$&$F{}_{0.25}^{0.35}$&$F{}_{0.35}^{0.45}$&\\
\hline
0.003&$4.206\cdot10^{-4}$&&$2.812\cdot10^4$&&1.584&0.974&0.547&\\
0.004&$3.302\cdot10^{-4}$&&$3.589\cdot10^4$&&1.584&0.974&0.547&\\
0.005&$3.318\cdot10^{-4}$&&$3.551\cdot10^4$&&1.584&0.974&0.547&\\
0.010&$3.685\cdot10^{-4}$&&$3.178\cdot10^4$&&1.589&0.974&0.545&\\
\end{tabular}
\caption{$\Delta Q$ dependence of the final results for a {\sc sw}\ system:
The
parameters and notation coincide with those of
tab.~\ref{tab:Qdrho:sw3:100}, except for $\Delta\rho = 5\cdot10^{-4} /
\sigma^3$.}\label{tab:Qdrho:sw3:2000}
\end{table}

Again, these predictions compare favorably with the numerical results
for a {\sc sw}\ potential of range $\lambda = 3$ obtained on the same
discretization grids as the {\sc hcy}\ data of
section~\ref{sec:smooth:rho:Q}.
The most prominent feature, barely compabitle with genuine smoothness,
{\it viz.{}}, near-discontinuities of $f$ can actually be found in the
numerical data underlying tabs.~\ref{tab:Qdrho:sw3:100}
and~\ref{tab:Qdrho:sw3:2000} at the locations marked with arrows in
fig.~\ref{fig:f:sw}; indeed, several of them can be seen clearly even on
the logarithmic scale of the graph.
All the other consequences of stiffness with ${Q_{\Delta Q}} > {Q_{\Delta
\rho}}$ are also
in agreement with the data of tabs.~\ref{tab:Qdrho:sw3:100}
and~\ref{tab:Qdrho:sw3:2000}: In particular, a pronounced $\Delta Q$
dependence of ${Q_{\rm min}}$ is accompanied by only a very modest effect
as
$\Delta \rho$ is varied, even though the relative change in $\Delta
\rho$ is much larger than that in $\Delta Q$.
Excluding the pathological data with negative $f$ ({\it v.~i.{}}), the
final
forms of $f$ are mostly $\Delta Q$ independent, and the forms obtained
on the two density grids differ but slightly.  Only the isotherm with
${\Delta Q\big\vert_\infty} = 0.010 / \sigma$ in
tab.~\ref{tab:Qdrho:sw3:2000}\ presents
a somewhat different shape than those at smaller ${\Delta
Q\big\vert_\infty}$.  The
differences in the numbers given under the heading $F^y_x$ are,
however, still in accordance with the stiff scenario as discrepancies
appear only close to the edge of the density range of large $f$.  As
for the first entry of tab.~\ref{tab:Qdrho:sw3:100}\ (${\Delta
Q\big\vert_\infty} =
0.003/\sigma$), negative $f$ corresponds to exceedingly small values
of $\varepsilon \equiv \bar\varepsilon + 1 \sim 10^{-27}$.  This is found
to be the
result of a downward jump from $f \sim +10^4$ ($\varepsilon \sim
10^{5000}$)
at only slightly higher cutoff where the form of $f$ again corresponds
to that of the other isotherms.  Clearly, even a minor perturbation of
the numerical process might easily have led to negative $\varepsilon$ and
hence to a numerical exception; in this case our implementation would
have discarded the last step, and the final results would once more
conform with those of the remainder of tab.~\ref{tab:Qdrho:sw3:100}.

Let us shortly return once more to the most salient
feature of the numerical solution, {\it viz.{}}, its near-discontinuities.
Disregarding the analytical considerations of part~I \cite{ar:10}\ it might
be tempting to imagine that, for $T <  T_c$, the
{\sc pde}\ generates a shock front approximately symmetrically moving
outward towards the densities $ \rho_v$ and $ \rho_l$ of the coexisting
phases as $Q$ approaches zero.  In this view of the numerical process
the jumps occur when the shock reaches the corresponding density.
Such an interpretation is not consistent with the data: According to
fig.~\ref{fig:f:sw}\ the near-discontinuities of $f$ occur repeatedly at
the same density (most conspicuously for $\rho = 0.1/\sigma^3$), and
rapid change at one density is generally accompanied by similar
behavior at other densities.  Neither of these observations is
compatible with the idea of a moving shock front, nor is there any
reason why the binodal should be linked to a shock front in {\sc sw}s\ but
not in the {\sc hcy}\ fluid, {\it cf.{}}\ section~\ref{sec:smooth:rho:Q}.

\subsection{Assertion of stiffness}

Summarizing the numerical evidence presented so far
we find that of the three scenarios found in part~I \cite{ar:10}\ only the
possibility of a merely logarithmic singularity of $f$ can be ruled
out with certainty.  We are then faced with the two alternatives of
genuine smoothness of the {\sc pde}\ on the one hand, and effective
smoothness as a
result of an {\sc fd}\ approximation to a stiff {\sc pde}\ on the other
hand.  As shown in
the preceding subsections~\ref{sec:smooth:rho:Q}
and~\ref{sec:smooth:Q:rho},
neither of them is in direct contradiction with the numerical data.

The crucial difference is their respective specificity and
testability: The genuinely smooth scenario does not make any
predictions beyond the smallness of the discretization grid dependence
of the numerical results, nor does it offer any of the detailed
understanding of the computational process that is necessary for
accurate and reliable interpretation of the {\sc fd}\ results.  By way of
contrast, stiffness of the {\sc pde}\ in part of its domain provides a
consistent framework for the interpretation of the numerics and
furthermore entails a number of concrete and numerically testable
consequences, all of which are in excellent agreement with our data
once the correct ordering of the ${Q_{\Delta x}}$ has been chosen.  In
combination with the analytical considerations of part~I \cite{ar:10}\ and
our
earlier statements regarding the importance of the formulation of the
{\sc hrt}\ {\sc pde}\ employed, the specificity and great number of these
predictions provide ample, although necessarily indirect evidence in
favor of the stiff scenario.

From this point on we will therefore take it for granted that the
{\sc hrt}\ {\sc pde}\ does, indeed, turn stiff in part of its domain for
subcritical temperatures.  On this basis we now aim to further enhance
our understanding of the {\sc hrt}\ numerics, shedding some light on the
location of ${Q_{\Delta Q}}$
(section~\ref{sec:heuristics}), extending our findings in the asymptotic
region to intermediate $Q$ (section~\ref{sec:beyond}), and finally
clarifying the outstanding numerical properties of the {\sc hcy}\ potential
{\it vis-\`a-vis} other physical systems (section~\ref{sec:hcy}).

\section{The onset of smoothing in $Q$}

\label{sec:heuristics}

Considering the great importance of the relative order of the
${Q_{\Delta x}}$ for the numerical process, it is natural to inquire into
their typical values.  As the exponents $r$ and $s$ are non-zero by
assumption, these cutoffs may only weakly depend on the discretization
grid and so are largely determined by the perturbational part of the
potential alone.  Whereas the onset of smoothing in $\rho$ eludes
simple reasoning so far, some heuristic arguemnts point to a simple
connection between the likelihood of finding ${Q_{\Delta Q}}$ at some
cutoff and the form of the
Fourier transform $\tilde w(k)$:

Let us consider a thermodynamic state of diverging isothermal
compressibility at a cutoff that is low enough for smoothing in $Q$ to
have set in at least partially, $Q \sim {Q_{\Delta Q}}$: In view of the
gradual
transition between the smoothing and non-smoothing regimes, the effective
exponent $ s_{\rm eff}$ may not
vanish exactly yet; nevertheless it seems safe to assume $ s_{\rm eff}<1$.
Of course we expect $\bar\varepsilon \gg f \gg 1$ so that reasoning based
on
the asymptotic behavior for large $\bar\varepsilon$ is applicable, and due
to
the monotonicity of the exponential function the likelihood of finding
${Q_{\Delta Q}}$ close to some cutoff $Q$ increases with the slope
$-\partial f / \partial Q$ of $f$.
At the same time, for a hard-sphere reference system ${Q_{\Delta Q}}$ can
only
depend on the form of the Fourier transform of the perturbational part
of the interaction potential, {\it i.~e.{}}, on $\tilde
u_0=\tilde\phi/\tilde\phi_0$ rather than
on $\tilde\phi$ itself: The temperature $T=1/k_B\,\beta$ enters the
calculation only as a pre-factor to the interaction potential, {\it
viz.{}},
through $\phi=-\beta\,w$ so that the normalization of $\tilde\phi$ only
fixes an energy or temperature scale.

With this in mind we define an auxiliary quantity $\psi(Q,\rho)$,
corresponding to $\tilde\phi_0+\gamma_0^{(Q)}$ in the notation of our
earlier work on
{\sc hrt}\ \cite{ar:10,ar:4,ar:th,ar:5}, through
\begin{equation} \label{def:psi}
\tilde{\cal K} + \psi\,\tilde u_0
=
- {\tilde\phi\over\bar\varepsilon}.
\end{equation}
Solving this relation for $\psi$ and differentiating with respect
to $Q$ we obtain
\begin{displaymath}
{\partial\psi\over\partial Q}
=
-\tilde\phi_0\,\left(
{\partial\over\partial Q}{1\over\bar\varepsilon}
+{\partial\over\partial Q}{\tilde{\cal K}\over\tilde\phi}
\right),
\end{displaymath}
which is valid at all cutoffs except close to the zeros $ Q_{\tilde\phi,i}$
of
$\tilde\phi$ and $\tilde u_0$ where eq.~(\ref{def:psi}) cannot be inverted.
--- An
alternative expression for $\partial\psi / \partial Q$ can be obtained
from the {\sc pde}~(\ref{pde:f}) and the compressibility sum rule:
Following
section~2.4.1 of ref.~\cite{ar:th}, for density independent potential we
easily find
\begin{displaymath}
\begin{array}{rl}
\displaystyle
{\partial\psi\over\partial Q}
=&\displaystyle
-{Q^2\over4\,\pi^2}\,{\partial^2\over\partial\rho^2}\ln\varepsilon
\\
=&\displaystyle
{Q^2\over4\,\pi^2}\,\left(
-\tilde u_0^2\,{\partial^2f\over\partial\rho^2}
+\tilde\phi\,{\partial^2\over\partial\rho^2}{1\over\tilde{\cal K}}
\right).
\end{array}
\end{displaymath}
Equating these two expressions for $\partial\psi/\partial Q$, solving
for $\partial^2f/\partial\rho^2$, and inserting the result into the
{\sc pde}~(\ref{pde:f}) yields
\begin{equation} \label{heur:dQf}
{\partial f\over\partial Q}
=
d_{00}
+ { d_{02}\,4\pi^2\over Q^2\,\tilde u_0^2} \,
\left(  \tilde\phi_0\,{\partial\over\partial Q}{1\over\bar\varepsilon}
+ {\partial\over\partial Q}{\tilde{\cal K}\over\tilde u_0}
\right)
+ { d_{02}\,\tilde\phi_0\over\tilde u_0} \,
{\partial^2\over\partial\rho^2}{1\over\tilde{\cal K}}
\end{equation}
for $Q$ away from the $ Q_{\tilde\phi,i}$.  Both $ d_{00}$ and $ d_{02}$
are negative in
the case under consideration \cite{ar:10}.

\begin{figure}[t]\vbox{\noindent\epsfbox{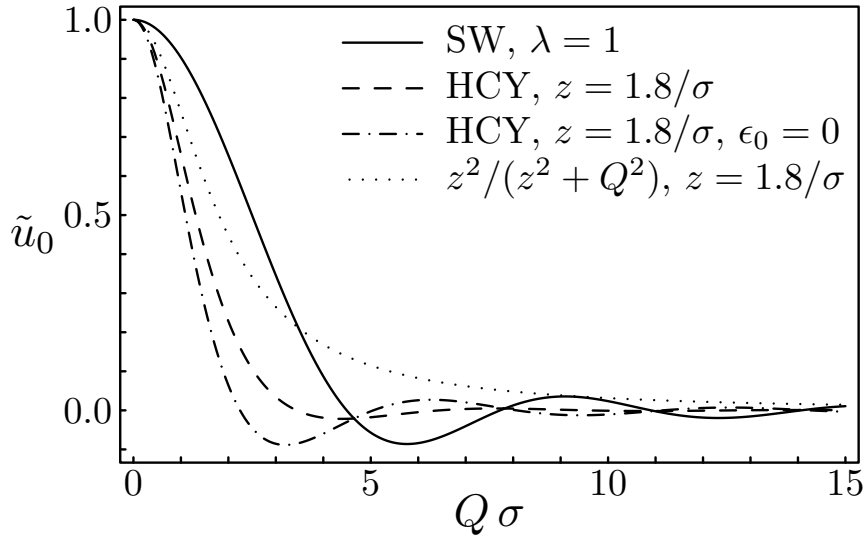}
\bigskip\caption{$\tilde u_0$ as a function of $Q$ for {\sc sw}\ and
{\sc hcy}\ potentials with the parameters indicated: If, contrary to
eq.~(\ref{pot:hcy}), the Yukawa form is retained even inside the core,
$\tilde u_0(Q)$ is given by $z^2/(z^2+Q^2)$.  As far as the {\sc sw}\
potential is concerned, $\lambda$ and $Q$ enter $\tilde u_0$ only in the
combination $\lambda\,Q$ so that a variation of the potential
range only introduces a linear rescaling of the $Q$ dependence of
the function. We have checked that the graph remains qualitatively
unchanged for different parameter settings.  The first minimum of
$\tilde u_0$ is $-0.02$ for the default {\sc hcy}\ potential, $-0.09$ for
the
{\sc hcy}\ potential with $\epsilon_0=0$, and slightly above $-0.09$
for the {\sc sw}\ potential.}\label{fig:heuristics}}
\end{figure}

Of the expressions appearing on the right hand side of eq.~(\ref{heur:dQf})
the one involving the $Q$ derivative of $1/\bar\varepsilon$ is of order
${\bf O}(\bar\varepsilon^{ s_{\rm eff}-1})$ and so can be neglected if $
s_{\rm eff}<1$ as assumed.  As we are looking for an effect
triggered by the form of $\tilde\phi$ alone we do not have to consider the
derivatives of the properties of the hard sphere reference system
encoded in $\tilde{\cal K}$ either.  It is then the term involving the $Q$
derivative of $\tilde u_0$ that is of interest: The ideal gas contribution
$-1/\rho$ to $\tilde{\cal K}$ ensures positive $ d_{02} \, \tilde{\cal K}$
so that this
term is the product of $\partial \tilde u_0 / \partial Q$ and manifestly
positive factors. Now assume that ${Q_{\Delta Q}}$ is less than the
position of
the first minimum of $\tilde u_0$ so that only the monotonous growth of
$\tilde u_0$ towards its global maximum at $Q=0$ remains to be covered by
the solution of the {\sc pde}: Clearly, as the calculation proceeds in the
negative $Q$ direction, the steeper this rise of $\tilde u_0$, the more the
$\partial \tilde u_0 / \partial Q$ term counteracts the growth of $f$,
thereby effectively further delaying the onset of smoothing in $Q$.
Most likely, ${Q_{\Delta Q}}$ will thus be found at cutoffs so low that
$\tilde u_0$
already levels off towards its limiting value of unity.

For the two potentials considered earlier, {\it viz.{}}\ {\sc sw}s\ and the
{\sc hcy}\ system with $z=1.8/\sigma$, fig.~\ref{fig:heuristics} shows that
$\tilde u_0$ levels off when $Q$
(or $\lambda\,Q$, in the case of {\sc sw}s) is no more than about
$10^{-1}/\sigma$, which is well compatible with the estimate of
section~\ref{sec:smooth:rho:Q}.  In addition, figs.~\ref{fig:f:hcy}
and~\ref{fig:f:sw} demonstrate that the
transition to the regime where $f$ mostly grows like $1/Q$,
corresponding to vanishing $ s_{\rm eff}$, occurs at similar values of the
cutoff. All in all, our arguments, heuristic as they are, do indeed allow
us to
estimate ${Q_{\Delta Q}}$ in a satisfactory way.  As for smoothing in
$\rho$,
on the other hand, actual numerical solution of the {\sc pde}\ currently is
the only
way of locating and studying ${Q_{\Delta \rho}}(T,\rho)$.

\section{Beyond asymptotics}

\label{sec:beyond}

\begin{figure}[t]\vbox{\noindent\epsfbox{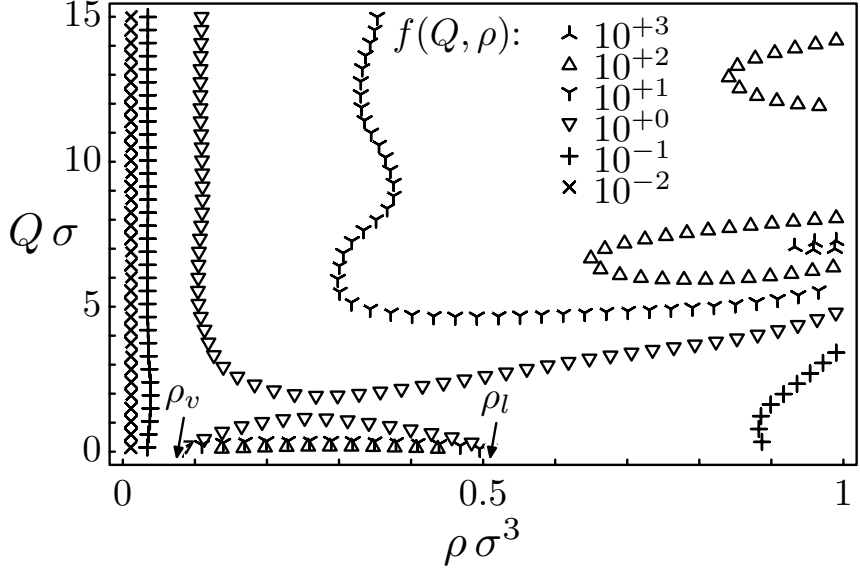}
\bigskip\caption{$f(Q,\rho)$ for intermediate cutoff as a
logarithmic contour plot: The data has been obtained for {\sc sw}s\
with $\lambda=3$ at inverse temperature $\beta = 1/k_B\,T =
0.115/\epsilon$.  Both the approach to the low-density boundary
condition of vanishing $f$ at densities below $0.01/\sigma^3$ and
the final build-up of infinite compressibility at cutoffs below
$10^{-1}/\sigma$ have been excluded from the graph.}\label{fig:f:contour}}
\end{figure}

On the basis of the results presented so far one might expect numerical
difficulties to first surface close to ${Q_{\Delta x}}$, {\it i.~e.{}},
around $Q \sim 10^{-1} / \sigma$ for the potentials considered
earlier. However, the monitoring variant of our code
\cite{ar:4,ar:th,ar:hrt:1} that must be credited with first highlighting
the stiffness
of the equations clearly signals the inadequacy of the discretization
grid already at much higher cutoff, {\it viz.{}}, typically for $5 <
Q\,\sigma
< 10$: Indeed, the asymptotic region of large $\bar\varepsilon$ can never
even
be reached without renouncing control of the local truncation error in
solving the {\sc fde}s, {\it cf.{}}\ section~III\,E of
ref.~\cite{ar:4}.

In combination with the observed patterns of the evolution of $f$ at
intermediate and small $Q$ illustrated in figs.~\ref{fig:f:hcy},
\ref{fig:f:sw},
and~\ref{fig:f:contour}, our experience with the numerics of {\sc hrt}\
leads us
to propose that stiffness is not confined to that part of ${\cal D}$ where
the
final build-up of infinite $\kappa_T$ takes place. Indeed,
fig.~\ref{fig:f:contour} shows that there are several regions of large $f$
at higher cutoff, some of which may give rise to transient stiffness
of the {\sc pde}: Even though the analysis of part~I \cite{ar:10} does not
apply
directly --- $f$ being bounded, asymptotic reasoning is not guaranteed
to be valid, nor does large $f$ imply large $\bar\varepsilon$ any longer
due
to the smallness of $\tilde u_0^2$ ---, from the expressions given in
part~I \cite{ar:10}\ we can still deduce that $ d_{02}$
is negative and appreciable for all $Q$ in the relevant cutoff range
except very close to the $ Q_{\tilde\phi,i}$,
and that $ d_{00}$ is likely to be rather large in modulus for $f \gg 1$
due to the terms linear in $f$.  Depending on the sign of $ d_{00}$, large
$f$ may well prompt rapid further growth when $Q$ proceeds to smaller
values.
Just as in
section~\ref{sec:smooth:Q:rho}, such a rapid growth of $f$ almost certainly
induces an accompanying growth of $\vert \partial^2 f / \partial \rho
^2 \vert$ on the grid, and any oscillations of the density curvature
will carry over to $\partial f / \partial Q$.  Qualitatively the
situation is then quite similar to that in the asymptotic region, and it
seems reasonable to see this transient stiffness
at intermediate cutoff as preventing computations insisting on local
convergence on a dynamically adjusted discretization mesh to ever
proceed to $Q \sim {Q_{\Delta x}}$.

Without the backing of more formal arguments much of the above line of
thought may seem insubstantial.  There are, however, a number of
numerical effects that provide at least indirect evidence for the
point of view just laid out.  Among those already discussed in our
earlier work on {\sc hrt}, the plummeting step sizes observed when
determining the discretization grid from the local curvature of
appropriate components of the solution vector \cite{ar:4,ar:hrt:1}
are the most direct sign of stiffness at intermediate $Q$.
Further support comes from our study of {\sc sw}s\ of varying range
\cite{ar:5}: There the peculiar shifts in the critical temperature
whenever $\lambda$ is close to a simple fraction have been linked to
the modulation of $\bar\varepsilon$ by the interference of $\tilde c_2^{\rm
ref}$ and
$\tilde\phi$; and considering our remarks on the effect of extending the
solution vector
(section~\ref{sec:numerics}) it is significant that the critical point is
accessible in a wider $\lambda$ range when coupling the {\sc pde}\ to a
smaller number of expansion terms for taking into account the core
condition, {\it cf.{}}\ section~IV\,E of ref.~\cite{ar:5} and appendix~E of
ref.~\cite{ar:th}.  Transient stiffness also explains why the lowest
temperature attainable numerically, denoted $1 / k_B \,  \beta_{{\rm
max},\tt\#}$
in refs.~\cite{ar:th,ar:5}, may well be higher than $ T_c$ even though
stiffness in the asymptotic region is a problem only for $T \le  T_c$,
and that the isotherms show no sign of phase separation for $\beta <
\beta_{{\rm max},\tt\#} <  \beta_c$, the critical temperature being known
independently from related computations or by other methods.  --- There are
also some more intricate issues related to
the interplay of the $ Q_{\tilde\phi,i}$ (where $ d_{00}$ vanishes as
$\tilde\phi^2$) with
the boundaries of the cutoff ranges where the step sizes $\Delta Q$
are inappropriate, as well as to the $\rho$ dependence of the onset of
smoothing in the presence of a local density grid refinement.
Discussion of these subtle effects and their numerical manifestations
requires a detailed presentation of appropriate methods of data
analysis on non-uniform high-resolution density grids and so falls
outside the scope of the present report.

\section{Hard core Yukawa {\it vs.} other potentials}

\label{sec:hcy}

In conjunction with our earlier analyses of the issues surrounding
initial and high density boundary conditions, implementation of the
core condition, and the peculiarities of discontinuous potentials
\cite{ar:4,ar:th,ar:5}, assertion of stiffness at low and intermediate
$Q$ below a certain temperature provides us with a detailed
understanding of the numerical process of solving the {\sc hrt}\ {\sc pde}\
throughout ${\cal D}$ and has proved invaluable in interpreting numerical
raw data.  We close this short series of reports with a generally
relevant sample of the kind of insight that can be gained on this
basis, {\it viz.{}}, a clarification of the unusually benign computational
properties of the {\sc hcy}\ potential:

Throughout our numerical work we consistently found that {\sc hcy}\ fluids
of moderate inverse screening length like, {\it e.~g.{}}, the one with $z =
1.8/\sigma$ repeatedly used here and in ref.~\cite{ar:4} exhibit the
symptoms of stiffness only in a rather mild form, both for $Q \to 0$
where this follows from the low value of ${Q_{\Delta Q}}$, and at
intermediate
cutoff.  This can be understood by noting, firstly, that the
temperature enters the calculation only through $\tilde\phi = - \beta\,w$,
the global normalization of which is accessible only in the limit $Q
\to 0$.  At any cutoff $Q$, the variation of $\tilde\phi(k)$ for $k > Q$ is
then the only measure of the temperature available to the {\sc pde}; at the
same time, for every one of the patches of large $f$, stiffness arises
only below some characteristic temperature, coinciding with the
critical one for the final build-up of infinite $\kappa_T$ at $Q=0$.
Secondly, recalling that $\tilde u_0 \propto \tilde\phi$, a look at
fig.~\ref{fig:heuristics} and the numbers quoted in its caption shows that
the local extrema at $Q > 0$ of $\tilde\phi^{{\rm hcy}}$ with the default
choice of
$\epsilon_0 = \epsilon$ are substantially smaller in modulus than
those for {\sc sw}s; only for much higher $z$ do the extrema of $\tilde
u_0^{{\rm hcy}}$
approach the {\sc sw}\ values which they reach in the infinite-$z$ limit.
It is easily checked that these observations also hold in comparison
with other short-ranged potentials like, {\it e.~g.{}}, the Lennard-Jones
one:
the main difference relative to {\sc sw}s\ concerns the phase rather than
the amplitude of the oscillations.  Taken together, the smallness of
the local extrema of $\tilde\phi^{{\rm hcy}}$ and the {\it r\^ole}\ the
temperature plays
readily explain the especially attractive numerical properties of this
potential: For $Q \to 0$, the slope of $\tilde u_0$ relative to the scale
set by the oscillations at higher $Q$ is particularly steep, as per
section~\ref{sec:smooth:rho:Q} leading to especially small ${Q_{\Delta Q}}$
and
suppression of near-discontinuities of $f$.  At intermediate cutoff,
on the other hand, the smallness of the local extrema {\it vis-\`a-vis} the
global
maximum at $Q = 0$ renders the numerics there similar to what would be
seen at much higher $T/T_c$ in other systems, and transient stiffness
poses less of a problem.  At the same time, the $z$-dependence of
$\tilde u_0^{{\rm hcy}}$ immediately explains the deteriorating accuracy of
the
results for very short {\sc hcy}\ screening length \cite{allg:7}.

Support for this view comes from the numerical properties following
from a different choice of $w(r)$ inside the hard core, which affects
the Fourier transform $\tilde\phi$ and hence $f$ and all other properties
of the $Q$ system at all cutoffs except in the limits $Q \to \infty$
and $Q \to 0$.  (Independence of the final results from the precise
choice of $w$ is confirmed in a rather satisfactory way in some
preliminary calculations on the Girifalco description of fullerenes
\cite{girifalco:1991}.)  The simplest such modification of $w$ consists
in a non-default setting of $\epsilon_0$ in eq.~(\ref{pot:hcy}),
exemplified by the dot-dashed curve in fig.~\ref{fig:heuristics}
($\epsilon_0=0$): Just as in ref.~\cite{ar:5}, even a modest discontinuity
of $w(r)$ at $r = \sigma$ strongly affects the form of $\tilde\phi$ and
renders the local extrema similar to those of the {\sc sw}\ case; as
expected, this is accompanied by numerical difficulties at
intermediate $Q$ similar to those discussed for {\sc sw}s\ in
refs.~\cite{ar:4,ar:th}.  In contrast, extension of the Yukawa form all the
way to the
origin --- hardly unproblematic as it entails diverging direct
correlation function at $r=0$ and invalidates the expansion method of
taking into account the core condition \cite{hrt:4} --- yields the
non-oscillatory form $\tilde u_0(Q) = z^2/(z^2+Q^2)$ (dotted curve in
fig.~\ref{fig:heuristics}) and prevents numerical solution of the {\sc
fde}s\
even at high temperatures.  The exceptionally attractive numerical
properties of the potential~(\ref{pot:hcy}) are therefore merely the result
of a particular choice, shared with the original implementation, of
$w(r)$ inside the core and so no genuine trait of the {\sc hcy}\ fluid.

This finding nevertheless does not invalidate the special standing of
the {\sc hcy}\ system that must be taken into account in interpreting a
comparison with other thermodynamically consistent liquid state
theories on the basis of {\sc hcy}\ results for $1.8 / \sigma \le z \le 9 /
\sigma$ \cite{allg:7}.  On the other hand, the above considerations also
point to
the possibility of tuning the computational properties of some given
potential by optimizing $w(r)$ inside the core to reduce the local
extrema at intermediate cutoff, an avenue largely unexplored to date
the merit of which we are currently in no position to assess.

The preceding clarification regarding the {\sc hcy}\ system {\it
vis-\`a-vis} other potentials is but one application of the detailed
understanding of the {\sc hrt}\ numerics presented here as well as in our
earlier {\sc hrt}\ related work.  Other aspects of the numerics where this
understanding has proved invaluable in interpreting the computational
process and the results it yields concern the limits of the resolution
in $\rho$ when using extremely fine density grids, the interplay
between non-uniform discretization grids and the location of the
binodal, the local behavior of the solution close to the zeros $
Q_{\tilde\phi,i}$
of $\tilde\phi$, or questions of data analysis.  All in all, we feel that
we have amassed a considerable amount of numerical experience and
arrived at a rather detailed self-consistent perception of the
computational process throughout all of ${\cal D}$ even below the critical
temperature.  Given the precarious nature of the {\sc hrt}\ numerics and
the not altogether unproblematic relation between the {\sc pde}\ and its
{\sc fd}\ approximation such an understanding is of prime importance if
systematic mistakes are not to be introduced into the results
unknowingly.

\section*{Acknowledgments}

The authors gratefully
acknowledge financial support from {\it Fonds zur F\"or\-der\-ung der
wissen\-schaft\-lichen Forschung} ({\it Austrian Science Fund}, FWF)
under projects~P14371-TPH, P15758-N08, and~J2380-N08.

\end{document}